\documentclass[twocolumn,tighten,times]{aastex62}

\hypersetup{linkcolor=black,citecolor=black,filecolor=cyan,urlcolor=magenta}

\accepted{for publication at ApJ}

\newcommand{\um}{$\mu$m}

\shorttitle{The far-infrared polarized arms of NGC~1068}
\shortauthors{Lopez-Rodriguez et al.}

\begin{document}

\title{SOFIA/HAWC+ traces the magnetic fields in NGC~1068}

\correspondingauthor{Lopez-Rodriguez, E.}
\email{enrique.lopez-rodriguez@nasa.gov}

\author{Enrique Lopez-Rodriguez}
\affil{SOFIA Science Center, NASA Ames Research Center, Moffett Field, CA 94035, USA}

\author{C. Darren Dowell}
\affil{NASA Jet Propulsion Laboratory, California Institute of Technology, 4800 Oak Grove Drive, Pasadena, CA 91109, USA}

\author{Terry J. Jones}
\affil{Department of Astronomy, University of Minnesota, 116 Church Street, SE, Minneapolis, MN55455, USA}

\author{Doyal A. Harper}
\affil{Department of Astronomy and Astrophysics, University of Chicago, Chicago, IL 60637, USA}

\author{Marc Berthoud}
\affil{Engineering + Technical Support Group, University of Chicago, Chicago, IL 60637, USA}
\affil{Center for Interdisciplinary Exploration and Research in Astrophysics (CIERA), Northwestern University, Evanston, IL 60208, USA}

\author{David Chuss}
\affil{Department of Physics, Villanova University, 800 E. Lancaster Ave., Villanova, PA 19085, USA}

\author{Daniel A. Dale}
\affil{Department of Physics \& Astronomy, University of Wyoming, Laramie, WY, USA}

\author[0000-0001-8819-9648]{Jordan A. Guerra}
\affil{Department of Physics, Villanova University, 800 E. Lancaster Ave., Villanova, PA 19085, USA}

\author[0000-0001-6350-2209]{Ryan T. Hamilton}
\affil{Lowell Observatory, 1400 W Mars Hill Rd, Flagstaff, AZ 86001, USA}

\author[0000-0002-4540-6587]{Leslie W. Looney}
\affil{Department of Astronomy, University of Illinois, 1002 West Green Street, Urbana, IL 61801, USA}

\author[0000-0003-3503-3446]{Joseph M. Michail}
\affil{Department of Astrophysics and Planetary Science, Villanova University, 800 E. Lancaster Ave., Villanova, PA 19085, USA}
\affil{Department of Physics, Villanova University, 800 E. Lancaster Ave., Villanova, PA 19085, USA}

\author{Robert Nikutta}
\affil{National Optical Astronomy Observatory, 950 N Cherry Ave, Tucson, AZ 85719, USA}

\author{Giles Novak}
\affil{Center for Interdisciplinary Exploration and Research in Astrophysics (CIERA), Northwestern University, Evanston, IL 60208, USA}
\affil{Northwestern University, Department of Physics and Astronomy, Evanston, IL 60208, USA}

\author[0000-0002-9650-3619]{Fabio P. Santos}
\affil{Max-Planck-Institute for Astronomy, K\"{o}nigstuhl 17, 69117 Heidelberg, Germany}

\author{Kartik Sheth}
\affil{NASA Headquarters, 300 E Street SW, DC  20546, USA}

\author{Javad Siah}
\affil{Department of Physics, Villanova University, 800 E. Lancaster Ave., Villanova, PA 19085, USA}

\author{Johannes Staguhn}
\affil{Dept. of Physics \& Astronomy, Johns Hopkins University, Baltimore, MD 21218, USA}
\affil{NASA Goddard Space Flight Center, Greenbelt, MD 20771, USA}

\author{Ian W. Stephens}
\affil{Harvard-Smithsonian Center for Astrophysics, 60 Garden Street, Cambridge, MA, USA}

\author[0000-0002-8831-2038]{Konstantinos Tassis}
\affil{Department of Physics and ITCP, University of Crete, Voutes, GR-70013 Heraklion, Greece}
\affil{Institute of Astrophysics, Foundation for Research and Technology-Hellas, 100 N. Plastira, Voutes, GR-70013 Heraklion, Greece}

\author{Christopher Q. Trinh}
\affil{USRA/SOFIA, NASA Armstrong Flight Research Center, Building 703, Palmdale, CA 93550, USA}

\author{Derek Ward-Thompson}
\affil{Jeremiah Horrocks Institute, University of Central Lancashire, Preston PR1 2HE, United Kingdom}

\author{Michael Werner}
\affil{NASA Jet Propulsion Laboratory, California Institute of Technology, 4800 Oak Grove Drive, Pasadena, CA 91109, USA}

\author[0000-0002-7567-4451]{Edward J. Wollack}
\affil{NASA Goddard Space Flight Center, Greenbelt, MD 20771, USA}

\author{Ellen G. Zweibel}
\affil{Department of Astronomy \& Physics, University of Wisconsin, Madison, WI 53706, USA}

\collaboration{(HAWC+ Science Team)}

\author{}



\begin{abstract} 
	We report the first detection of galactic spiral structure by means of thermal emission from magnetically aligned dust grains. Our 89 \um\ polarimetric imaging of NGC 1068 with the High-resolution Airborne Wideband Camera/Polarimeter (HAWC+) on NASA’s Stratospheric Observatory for Infrared Astronomy (SOFIA) also sheds light on magnetic field structure in the vicinity of the galaxy’s inner-bar and active galactic nucleus (AGN).
	We find correlations between the 89 \um\ magnetic field vectors and other tracers of spiral arms, and a symmetric polarization pattern as a function of the azimuthal angle arising from the projection and inclination of the disk field component in the plane of the sky. 
	The observations can be fit with a logarithmic spiral model with pitch angle of $16.9^{+2.7}_{-2.8}$$^{\circ}$ and a disk inclination of $48\pm2^{\circ}$.
	We infer that the bulk of the interstellar medium from which the polarized dust emission originates is threaded by a magnetic field that closely follows the spiral arms. Inside the central starburst disk ($<1.6$ kpc), the degree of polarization is found to be lower than for far-infrared sources in the Milky Way, and has minima at the locations of most intense star formation near the outer ends of the inner-bar. 
	Inside the starburst ring, the field direction deviates from the model, becoming more radial along the leading edges of the inner-bar. The polarized flux and dust temperature peak $\sim 3-6\arcsec$ NE of the AGN at the location of a bow shock between the AGN outflow and the surrounding interstellar medium, but the AGN itself is weakly polarized ($< 1$\%) at both 53 and 89 \um.

\end{abstract} 




\section{Introduction} \label{sec:int}

Over the past few decades, astronomers have detected the presence of magnetic fields in galaxies at all spatial scales. These major studies have been performed using optical and radio observations \citep[][for reviews]{kron1994,ZH1997,BG2004,beck2015}. Radio observations measure the polarization of synchrotron radiation from relativistic electrons; this radiation is sensitive to the cosmic ray electron population, which does not closely trace the gas mass. Studies of interstellar polarization at optical wavelengths can reveal the magnetic field geometry as a result of magnetically-aligned dust grains by radiative alignment torques \citep{ALV2015}. However, the optical polarization measurements suffer from contamination by highly polarized scattered light. Linear polarization at 6.3 and 21.2 cm using the Effelsberg 100-m telescope and optical polarization \citep{scar87} have been compared in the face-on spiral galaxy M~51 \citep{BKW1987,flecher2011}. The optical and radio linear polarization shows a similar magnetic field morphology within the eastern and southern quadrants, but with significant variations of the magnetic field direction, up to 60$^{\circ}$, in the western quadrant. A study on the face-on spiral galaxy NGC 6946 also show similarities between the optical and radio polarization \citep{FBN1998}.  Further observations of M~51 using near-infrared (NIR) polarization only registered upper-limit polarization across the galaxy, which ruled out the dichroic absorption as the main polarization mechanism \citep{pave12}. The scattering cross section of typical interstellar dust declines much faster $(\sim \lambda^{-4})$ between $0.55$ and $1.65~\micron$ than its absorption \citep[$\sim \lambda^{-1}$,][]{jowh15}. It is likely that the optical polarization measured in the previous works \citep[i.e.][]{scar91} was due to scattering, rather than extinction by dust grains aligned with the interstellar magnetic field. For M51, the expected dichroic absorptive polarization at H-band, based on the measured optical polarization, is $\sim0.4$\%, but an upper limit of $0.05\%$ was measured at H-band \citep{pave12}. The dichroic absorptive polarization should has shown a higher polarization at H-band than the measured to be due to magnetically aligned dust grains. Observations of the magnetic field geometry more sensitive to the denser gas and dust are needed.

Our understanding of how spiral arms form and their role in galaxy evolution is still incomplete. The most widely accepted theoretical model for spiral arms is the density-wave theory \citep{lindblad1960,LS1964,shu2016}. This theory posits that spiral structure can be described as a superposition of waves of enhanced stellar density with constant pitch angles (the angle between the tangent of the wave and circles around the galactic center) and constant pattern speed \citep{Athanassoula1984}. This theory predicts that stars form in the arms as gas moves into the wave and is compressed by its gravitational potential. Under this scheme, the spatial displacement of the spiral arms should be different for different tracers of star formation (e.g., molecular clouds, HII regions, and newly formed stars) because they appear at different phases of the wave. The spiral arms at optical/NIR wavelengths is expected to trace already born stars, while far-IR (FIR) wavelengths will trace ongoing star formation.

NGC 1068 is the nearest (D$_{L} = 13.5$ Mpc, 1\arcsec = 65~pc) grand-design spiral galaxy with both a bright active galactic nucleus (AGN) and a luminous circumnuclear starburst. This galaxy is classified as a Seyfert 2, where the active nucleus is obscured by a dusty structure, and the host is an Sb type. Associated with the AGN is a narrow-line region, i.e. ionization cones, of $\sim20\arcsec$ ($\sim1.3$~kpc) in diameter at a position angle of $\sim40^{\circ}$ East of North \citep[i.e.][]{BH1985} with the northern region protruding toward us out of the plane of the galaxy. \citet{Schinnerer2000} have presented CO observations of the central region of NGC 1068 and discussed the relationship between the stellar and gas dynamics in the galaxy. At radii $>15\arcsec$ ($>0.975$ kpc), the kinematic axis of the galaxy is approximately east-west, and at very large radii the eccentricities of the galaxy's faintest contours is consistent with the  $40^{\circ}$ inclination suggested by HI data. However, there is a bright oval structure $\sim180\arcsec$ ($\sim11.7$ kpc) diameter with long axis perpendicular to the kinematic axis that probably corresponds to a large-scale stellar bar \citep[fig. 6 of][]{Schinnerer2000}. The inner Lindblad resonance (ILR) of this bar has a de-projected radius of $\sim18\arcsec$ ($\sim1.17$ kpc), near the location of a compact spiral or ring-like structure in $^{12}$CO, consistent with theoretical predictions that gas can be transported inward along bars and accumulate at the ILR. NIR observations in the K-band have detected the presence of a $30\arcsec$ ($1.95$ kpc) diameter inner-bar \citep{scoville1988,thronson1989}. \citet{TD1988} have suggested that for the case of the NGC~1068 outer-bar the ILR is actually split into two resonances (``inner-inner''  or iILR and ``outer-inner'' or oILR) and that the CO spiral pattern is produced by a spiral density wave between the iILR and oILR that is driven by the inner-bar. We will refer to this annulus as the ``starburst ring''. In addition to the CO in the starburst ring, there are CO clumps extending down into the northeastern branch of the inner-bar (but not in the southwestern branch) and a compact CO ring at radii $\sim1\arcsec$ ($\sim 65$ pc) that may be gas accumulating at the ILR of the inner-bar.

There is ample evidence for an exceptionally high rate of star formation in the central region of NGC 1068, possibly the result of  a recent minor merger \citep{TYT2017}. The optical surface brightness of the galactic disk at radii $\le 25\arcsec$ ($\le1.63$ kpc)  (henceforth the ``starburst disk'') is among the brightest of any galaxy in the local universe \citep{KW1978,weedman1985,IOK1987}. Observations in the FIR have shown that the total luminosity of the starburst disk is $> 10^{11} L_{\odot}$ \citep{TH1980,telesco1984}. Radio observations at 1.465 GHz \citep{WU1982} show that the same region is extremely rich in the type of non-thermal synchrotron emission produced in supernova explosions of massive young stars. On the basis of 10.8 \um\ and NIR data, \citet{TD1988} and \citet{thronson1989}, respectively, have argued that almost all the actual star formation takes place within the starburst ring and that the most intense activity occurs near the outer ends of the inner-bar.

These properties have made NGC 1068 a suitable object for spatially resolved polarimetry. Using optical polarimetry of NGC~1068, \cite{scar91} found a spiral pattern that was interpreted as delineating the magnetic field geometry in the spiral arms of the galaxy. Given the effects of scattering on optical polarimetry measurements and that the host galaxy of NGC 1068 has not been observed using radio polarimetry, this association has to be questioned. 

We have performed FIR polarimetric observations to image the central $2\arcmin \times 2\arcmin$ ($7.8 \times 7.8$ kpc$^{2}$) of NGC~1068 at wavelengths of 53 and 89 \um. We discuss the 53 \um\ flux from the AGN in a previous paper \citep{LR2018a}. In this paper, we present our polarimetric results, including an 89 \um\ image that for the first time reveals galactic spiral structure by means of thermal emission from magnetically aligned dust grains. At FIR wavelengths, scattering and Faraday rotation are not a factor at these scales. The dominant emission is from warm dust, which more closely samples the total gas column density than relativistic electrons producing the synchrotron emission. The paper is organized as follows: Section \ref{sec:obs} describes the observations, data reduction, and observational results. In Section \ref{sec:Bfield}, we fit our polarimetry data to a spiral galactic magnetic field model, which is then analyzed and discussed in Section \ref{sec:DIS}. In Section \ref{sec:CON} we present our conclusions.


\section{Observations and Data Reduction} \label{sec:obs}

\begin{figure*}[ht!]
\includegraphics[angle=0,scale=0.6]{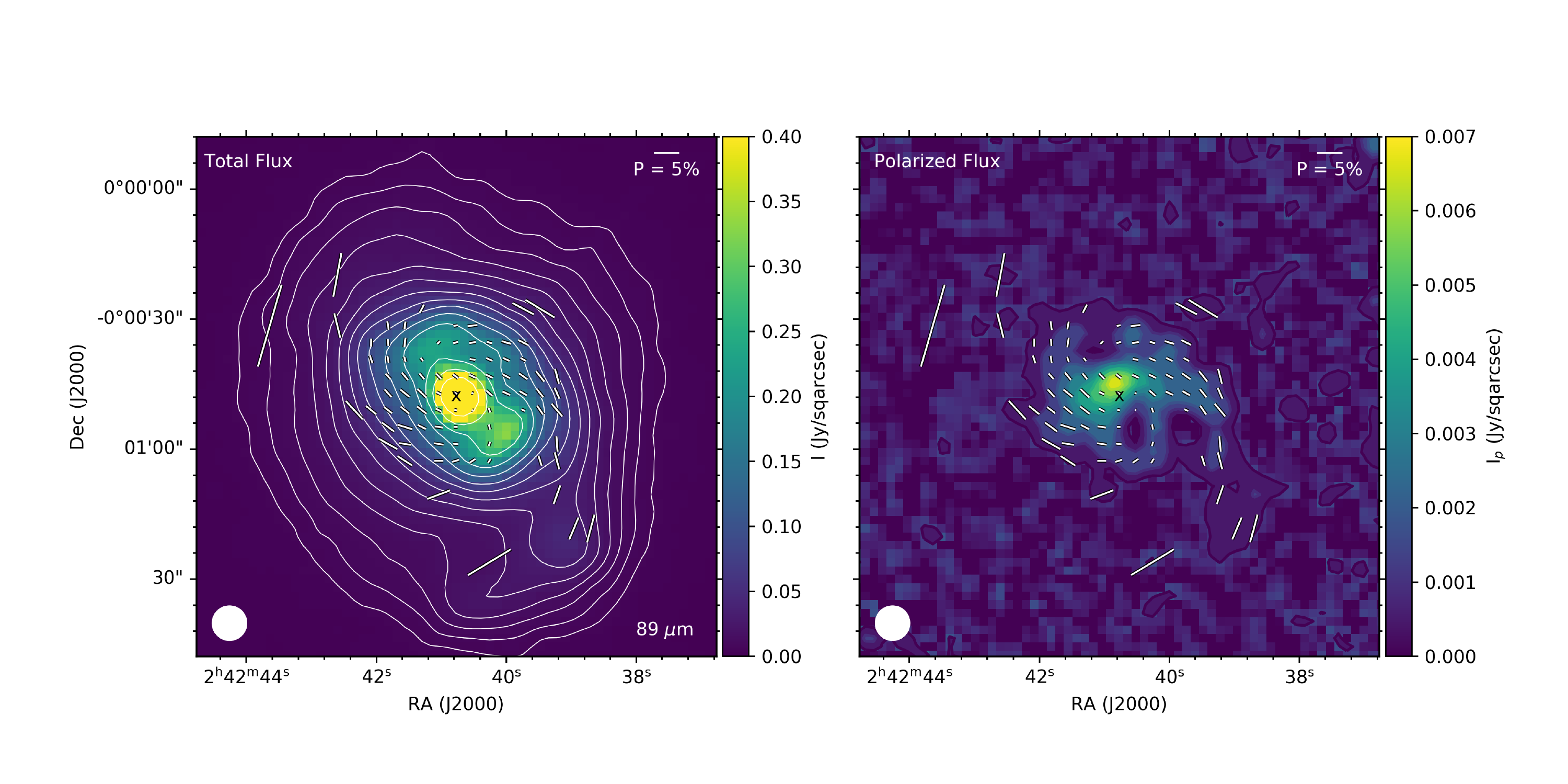}
\caption{\textit{Left:} 89 \um\ total flux (color scale) with polarization vectors (white vectors) with $P/\sigma_{P} > 3.0$  and rotated by $90^{\circ}$ to show the inferred magnetic field within the central $2\arcmin\ \times 2\arcmin$ ($7.8 \times 7.8$ kpc$^{2}$). Contours are shown for $2^{n}\sigma$, where $n = 3.0, 3.5, 4.0, \dots$ and with $\sigma = 6.74\times 10^{-3}$ Jy sqarcsec$^{-1}$. Polarization vector of $5$\% (white vector) and beam size of $7$\farcs$8$ (white circle) are shown. \textit{Right:} Polarized flux (color scale) with filled contours starting at $3\sigma$ with $\sigma = 6.64\times10^{-4}$ Jy sqarcsec$^{-1}$ and increasing in steps of $2\sigma$. Polarization vectors are shown as in the left figure. Black cross shows the location of the AGN. 
\label{fig:fig1}}
\epsscale{2.}
\end{figure*}

\begin{figure*}[ht!]
\includegraphics[angle=0,scale=0.5]{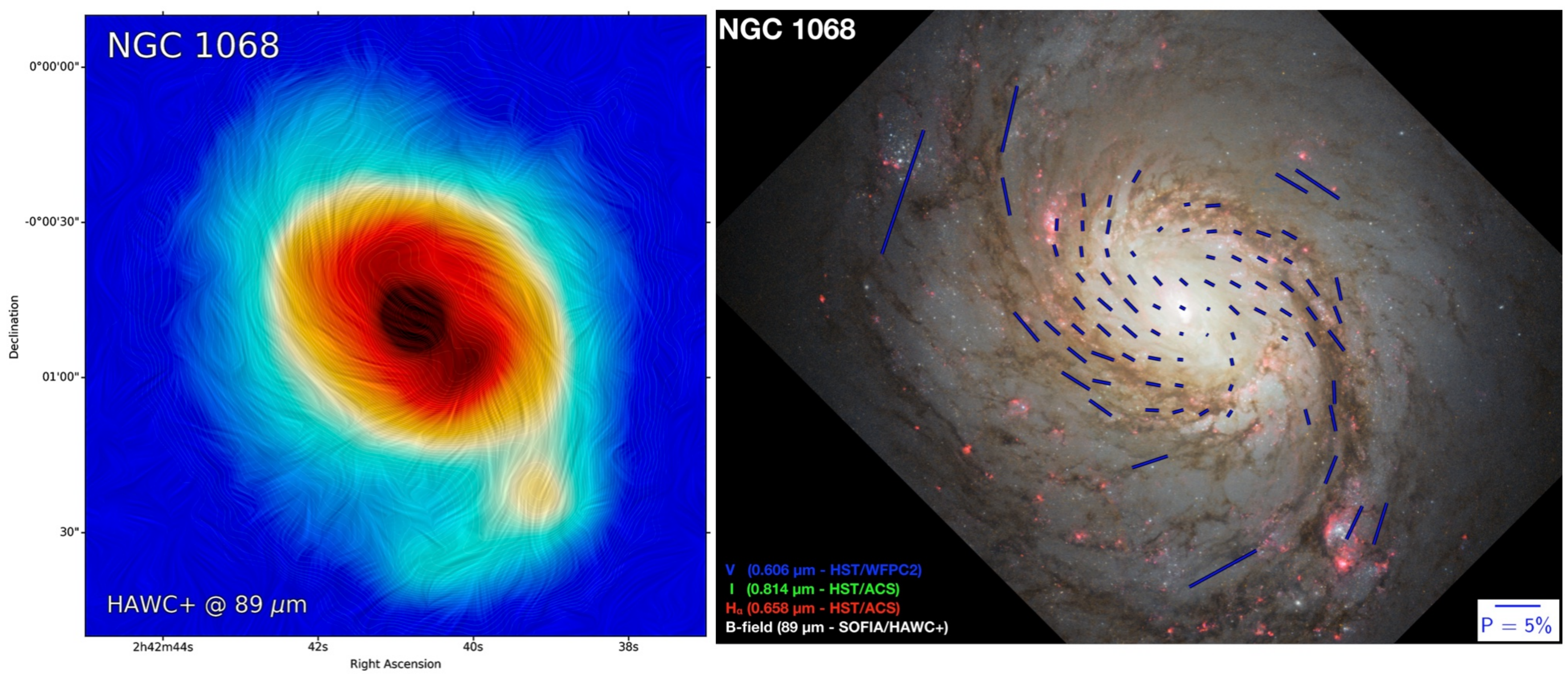}
\caption{\textit{Left:} Total flux (color scale) image at 89 \um\ with overlaid streamlines of the inferred magnetic field morphology using the line integral convolution technique. \textit{Right:} \textit{HST} composite image with overlaid magnetic field vectors (blue vectors) inferred by the SOFIA/HAWC+ 89 \um\ polarimetric observations. Polarization vectors have been rotated by $90^{\circ}$ to show the inferred magnetic field. Polarization vector of $5$\% (blue vector) is shown. \label{fig:fig2}}
\epsscale{2.}
\end{figure*}

NGC~1068 was observed at $53$ and $89$ \um~as part of the guaranteed time observations (GTO, ID: 70\_0609) using the High-resolution Airborne Wideband Camera-plus (HAWC+) 
\citep{Vaillancourt2007,Dowell2010,Harper2018} on the $2.7$-m Stratospheric Observatory For Infrared Astronomy (SOFIA) telescope. HAWC+ polarimetric observations simultaneously measure two orthogonal components of linear polarization arranged in two arrays of $32 \times 40$ pixels each, with pixel scales of $2$\farcs$55$ and $4$\farcs$02$ pixel$^{-1}$, and beam sizes (full width at half maximum, FWHM) of $4$\farcs$85$ and $7$\farcs$80$ at $53$ and $89$ \um, respectively. For both bands, we performed observations in a four-dither square pattern with a distance of three pixels from the center position in the array coordinate system. In each dither position, four halfwave plate position angles (PA) were taken. We used a chop frequency of 10.2 Hz, a nod time of $45$s, a chop-throw of $180$\arcsec, and a chop-angle of $90^{\circ}$ to always be along the short axis of the $32 \times 40$ pixels array. We used the science instrument reference frame (SIRF) for these observations. At $53$ \um, we observed a total of $8$ dither positions, and a total of $63$ dither positions at $89$ \um. The total on-source integration times were $0.47$h and $2.03$h at $53$ and $89$ \um, respectively. The total clock time of the observations—on-source time plus off-source time plus overheads—is 1.3h and 5.2h, respectively. Based on the morphology of the source using images from the \textit{Herschel Space Observatory}\footnote{\textit{Herschel} is an ESA space observatory with science instruments provided by European-led Principal Investigator consortia and with important participation from NASA} \citep{Pilbratt2010} and the chop-throw of $180\arcsec$ used for the HAWC+ observations, the reference beam fluxes should be $\le 1/10$ times the source flux over a source diameter of $120\arcsec$ and therefore negligible. Data were reduced using the \textsc{hawc\_dpr pipeline} v1.3.0. The pipeline procedure described by \citet{Harper2018} was used to background-subtract and flux-calibrate the data and compute Stokes parameters and their uncertainties. Computation of final degree and PA of polarization accounts for correction of the instrumental polarization \citep[table 4 by][]{Harper2018} with typical standard deviations after subtraction of $\sim0.3\%$, and de-biasing and PA error estimation \citep{WK1974}. Further analyses and high-level displays were done with custom \textsc{python} routines.

Figure \ref{fig:fig1} shows the total flux image with overlaid polarization vectors in a $2\arcmin \times 2\arcmin$ ($7.8 \times 7.8$ kpc$^{2}$) field-of-view (FOV) at $89$ \um. Polarization vectors have been rotated by $90^{\circ}$ to show the morphology of the magnetic field. We note that the polarization orientations are ambiguous by $180^{\circ}$, thus the displayed polarization vectors for all figures should be interpreted as magnetic field lines. Each polarization vector shows a statistically independent polarization measurement with $P/\sigma_{P} \ge 3$. Polarization vectors are shown every $4.02\arcsec$ (Nyqvist sampling at 89 \um). Throughout the paper we assume that dust grain alignment is perpendicular to the direction of the magnetic field. We observe extended dust emission along the spiral arms of the galaxy up to a diameter of $\sim$100\arcsec\ ($\sim6.5$~kpc) centered at the nucleus while the central $\sim1$ kpc radius emission is extended along PA of $\sim45^{\circ}$ and co-spatial with the inner-bar. The polarization map shows a spiral shape with a diameter of $\sim50$\arcsec\ ($\sim 3.25$ kpc). This spiral shape can be clearly identified in Figure \ref{fig:fig2}-left in the image made using the line integral convolution contour (LIC) technique \citep{CL1993}. This figure uses all polarization vectors regardless of signal-to-noise ratio measured in our observations. Figure \ref{fig:fig2}-right shows combined \textit{HST} observations at V-band, I-band, and H$_{\alpha}$ with overlaid magnetic field vectors (blue vectors) from our 89 \um\ HAWC+ observations. The polarized flux is shown in Figure \ref{fig:fig1}-right. The polarized flux shows extended emission along the E-W direction with a peak  shifted by $4\arcsec$ (Section \ref{sec:AGN}) NE from the peak of the total flux intensity.

Given the short integration time at $53$ \um, only polarization at the nucleus (one independent vector is measured) of NGC~1068 was detected, and the polarization map is not shown. For the total intensity image at 53 \um, we refer to the HAWC+ observations using the Lissajous observing mode presented by \citet{LR2018a}. We here report the measured nuclear degree of polarization of $1.3\pm0.3$\% and PA of polarization of $139\pm6^{\circ}$ (E-vector) within a $5$\arcsec\ ($325$ pc) diameter at $53$ \um. For 89 \um, we measured a nuclear polarization of $0.6\pm0.3$\% and PA of polarization of $156\pm13^{\circ}$ (E-vector) within a $8\arcsec$ ($520$ pc) diameter. For both bands, the measured nuclear polarization is below the instrumental polarization and above the residual polarization after instrumental polarization correction. Thus, the measured polarization is consistent with a low polarized core. As the fractional contribution of the AGN increases at short wavelengths and the dust emission from the host galaxy increases at long wavelengths \citep[fig. 6 by][]{LR2018a}, we expect to measure a decrease of the nuclear degree of polarization as wavelength increases. Our measurements are compatible with this behavior.


\section{Model of the large-scale galactic magnetic field}\label{sec:Bfield}

The observational results shown in Section \ref{sec:obs} suggest the presence of a large-scale magnetic field along two spiral arms. We here produce a model of the magnetic field  to characterize this structure. We use an axisymmetric spiral structure \citep[i.e.][]{RG2010,BHB2010} defined in cartesian coordinates by the form

\begin{eqnarray}
B_{x} &=& -B_{0}(r) \sin (\phi + \Psi)\cos \chi(z) \\
B_{y} &=& B_{0}(r) \cos (\phi + \Psi)\cos \chi(z)  \\ 
B_{z}   &=& B_{0}(r) \sin \chi(z)
\end{eqnarray}
\noindent
where $B_{0}(r)$ is the magnetic field as a function of the radial distance from the core of the galaxy, and $\Psi$ is the pitch angle defined as the angle between the azimuthal ($\phi$) direction and the magnetic field direction. $\chi(z)$ is the out-of-plane magnetic field taken as $\chi(z) = \chi_{0}\tanh(z/z_{0})$ \citep{RG2010}.

When this magnetic field configuration is viewed at an inclination, $i$, and tilt angle\footnote{Also described as the PA of the major axis of the projected galaxy plane.}, $\theta$, of the galaxy projected on the plane of the sky, the magnetic field at the observer's frame is given by

\begin{eqnarray}
B_{x^{s}} &=& B_{x}\cos\theta + (B_{y}\cos i - B_{z}\sin i)\sin\theta \\
B_{y^{s}} &=&  -B_{x}\sin\theta + (B_{y}\cos i - B_{z}\sin i) \cos\theta \\
B_{z^{s}} &=& B_{y}\sin i + B_{z}\cos i
\end{eqnarray}
\noindent
where ($x^{s},y^{s},z^{s}$) are the major axis, minor axis, and the line of sight, respectively, in the sky coordinate system. 

The PA of polarization in the plane of the sky, which is parallel to the direction of the magnetic field in the ($x^{s},y^{s},z^{s}$) coordinate system, is computed as $PA_B = \arctan(B_{y^{s}} / B_{x^{s}})$. The degree of polarization as a function of the inclination and magnetic field direction in the plane of the sky is computed as $P = p_{o} (1 - (B_{z^{s}}/B_{s})^{2})$, where $p_{o}$ is a constant factor, and $B^{2}_{s} = B^{2}_{x^{s}} + B^{2}_{y^{s}} + B^{2}_{z^{s}}$ is the total magnetic field. We assume a negligible contribution of the out-of-plane magnetic field, $\chi_{0} = 0$, thus our final magnetic field is coplanar with the disk of the galaxy. This model is purely a geometrical description and does not fulfill the divergence-free vector field as $B_z$ is neglected.

Since the FIR polarimetry is not directly sensitive to the magnetic field strength, there are three free model parameters: pitch angle, $\Psi$, inclination angle, $i$, and tilt axis position angle, $\theta$. Face-on view corresponds to $i = 0^{\circ}$ and edge-on to $i = 90^{\circ}$. Tilt axis has a reference point, $\theta = 0$, along the north-south direction and positively increases east from north. To fit for model parameters, we computed a Markov Chain Monte Carlo (MCMC) approach using the differential evolution metropolis sampling step in the \textsc{python} code \textsc{PyMC3} \citep{pymc}. The prior distributions are set to flat within the ranges $\Psi$~=~($0^{\circ}$,$50^{\circ}$), $i$~=~($0^{\circ}$,$90^{\circ}$), and $\theta$~=~($0^{\circ}$,$90^{\circ}$). We run the code using 15 chains with 6000 steps and a 1000 steps burn-in per chain. Because the central $8\farcs3 \times 8\farcs3$ ($0.54 \times 0.54$ kpc$^{2}$) region of NGC~1068 is dominated by the inner-bar, we have excluded this region in this analysis.  One could perhaps question the size of the exclusion zone, but the overall effect on the model fit (presented above) of the small number of vectors within the innermost annulus will be small.

\begin{figure}[ht]
\includegraphics[angle=0,scale=0.51]{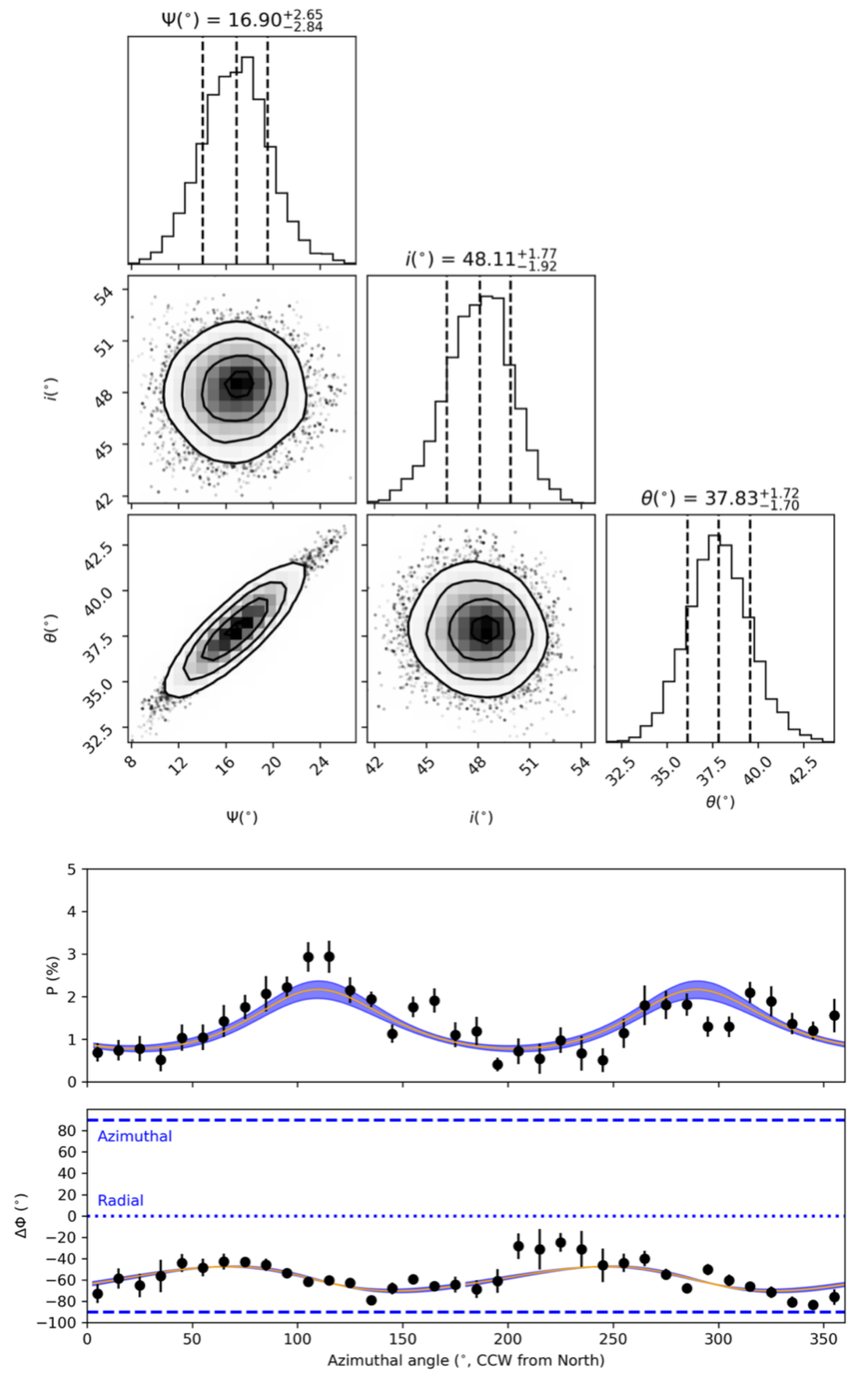}
\caption{\textit{Top:} Posterior distributions and MAP values of the pitch, $\Psi$, inclination, $i$, and tilt, $\theta$, angles of the magnetic field model. \textit{Bottom:} Degree and PA of polarization vectors were subtracted by vectors in radial coordinates centered on the galaxy nucleus, $\Delta\phi$ (black dots). Azimuthal angle equal to 0 corresponds to North and positive values are counted counter-clockwise in the East of North direction. $\Delta \phi = 0^{\circ}$ corresponds to a perfectly radial direction, while $\Delta \phi = \pm90^{\circ}$ corresponds to a perfectly azimuthal direction. The best fit model (orange solid line) and the $1\sigma$ uncertainties (blue shadow region) are shown.
\label{fig:fig3}}
\epsscale{2.}
\end{figure}

\begin{figure*}[ht!]
\includegraphics[angle=0,scale=0.28]{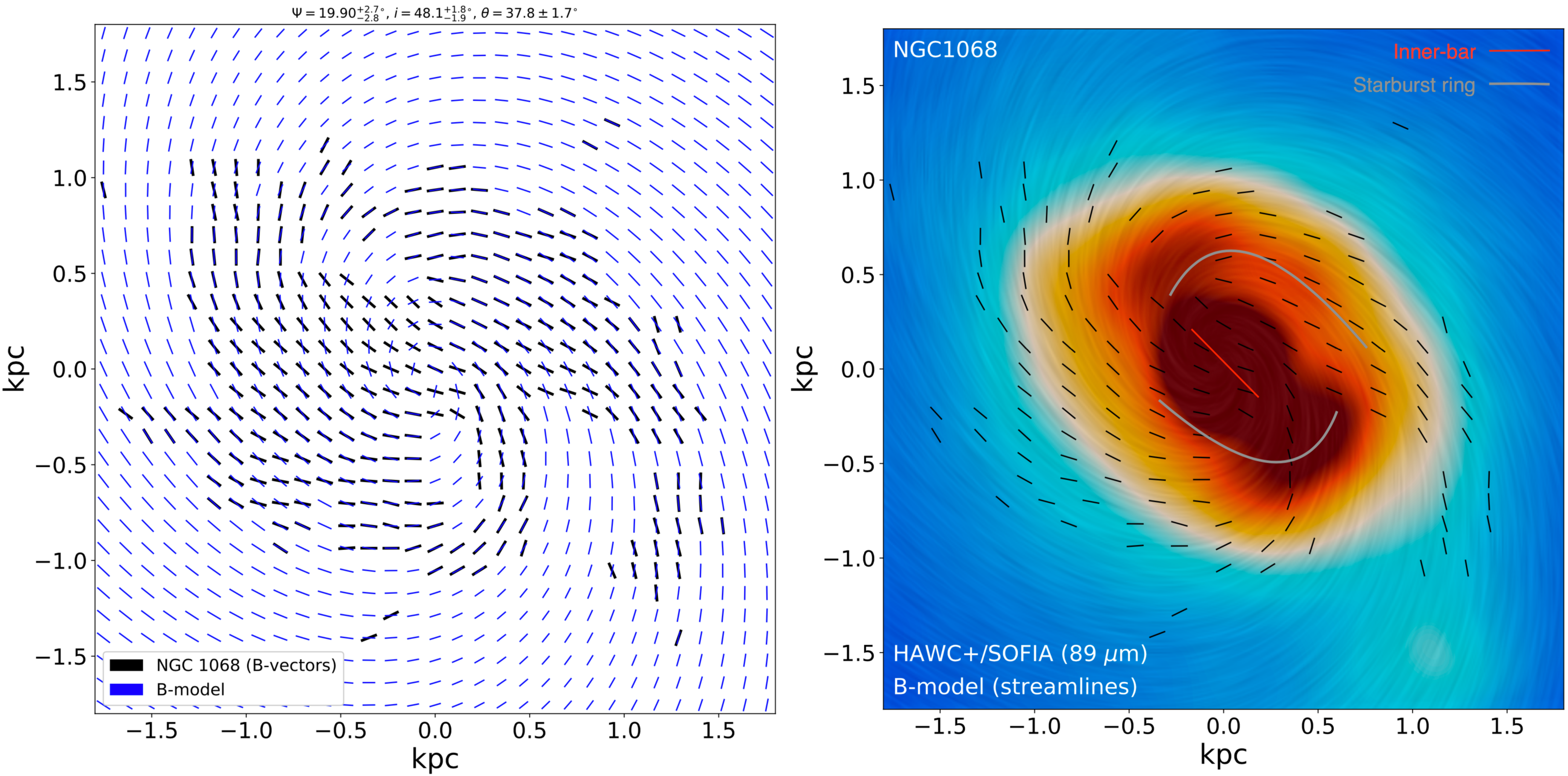}
\caption{\textit{Left:} Best inferred model (blue lines) of the galactic magnetic field explaining the measured (black lines) magnetized spiral arms of the central $3 \times 3$ kpc$^{2}$ of NGC1068. This figure shows constant length observed vectors (black) and predicted model vectors (blue). \textit{Right:} Total flux (color scale) with overlaid magnetic field model (streamlines) using LIC and measured polarization vectors (black lines). Approximated locations of the inner-bar and starburst rings are shown (Section \ref{sec:AGN}). Polarization vectors have been rotated by $90^{\circ}$ to show the inferred magnetic field.
\label{fig:fig4}}
\epsscale{2.}
\end{figure*}

Figure \ref{fig:fig3} shows the marginalized posterior distributions of the free parameters and their maximum a posteriori (MAP) with their 1$\sigma$ estimations. This figure also shows, as a function of azimuthal angle, the measured degree of polarization, the measured PA of polarization $\Delta\phi$ (black dots) in radial coordinates centered on the galaxy nucleus, the best fit model (orange solid lines), and 1$\sigma$ uncertainties (blue shadow region). If $\Delta \phi = 0^{\circ}$, the angle is perfectly radial, and if $\Delta \phi = \pm90^{\circ}$, the angle is perfectly azimuthal. To generate the data points in the figure, the Stokes I, Q, and U images were binned by azimuthal angle in bins of 10$^{\circ}$. Then, debiased degree and PA of polarization were estimated for each bin. 

We find a symmetric pattern of the polarization as a function of the azimuthal angle, reflecting the relationship among degree of polarization, polarization PA, and model parameters shown above. An axisymmetric spiral structure with a pitch angle of $\Psi = 16.9^{+2.7}_{-2.8}$$^{\circ}$, inclination $ i = 48.1^{+1.8}_{-1.9}$$^{\circ}$, and tilt angle of $\theta = 37.8\pm1.7$$^{\circ}$ best describes our data. The host galaxy is inclined by 40$^{\circ}$ \citep{deVaucouleurs1991}, and using the HI velocity field, \citet{brinks1997} estimated an inclination angle of $40\pm3^{\circ}$, which both are fairly close to our estimated values. The main difference may be that \citet{deVaucouleurs1991} estimated the galaxy inclinations using isophotes on the optical images, and that \citet{brinks1997} used the rotation curve of the HI emission observations while our estimation is based on a smaller region of the galaxy dominated by the FIR emission. The position angle of the kinematic major axis of HI is estimated to vary from  $270\pm3^{\circ}$ in the inner $30-70\arcsec$ disk to $286\pm5^{\circ}$ beyond $100\arcsec$, which is almost East-West \citep{brinks1997}. We further discuss the pitch angle in Section~\ref{sec:PA}.

Figure \ref{fig:fig4} (left) shows the magnetic field configuration (blue vectors) of the best inferred parameters (Fig. \ref{fig:fig3}) with the overlaid measured PA of polarization rotated by $90^{\circ}$ (black vectors) of the central $3 \times 3$ kpc$^{2}$ of NGC~1068. Figure \ref{fig:fig4} (right) shows the total flux (color scale) image of NGC~1068 with overlaid magnetic field model (streamlines) using LIC techniques and the measured polarization vectors (black lines). All polarization vectors have the same length, and their orientation shows the magnetic field morphology.  The large-scale magnetic field morphology agrees with an axisymmetric spiral structure. The largest difference between our model and the observations is located within $\sim 1$ kpc of the nucleus (Section \ref{sec:AGN}). As our axisymmetric magnetic field model does not fulfill the divergence-free requirement, we cannot reproduce the field in the inner areas of the galaxy. The deviations within $\sim 1$ kpc around the nucleus may indicate a limitation of the model and/or a different magnetic field morphology as described in Section \ref{sec:AGN}.


\section{Discussion}\label{sec:DIS}

\subsection{Temperature and column density maps}\label{sec:TNh}

\begin{figure}[h!]
\includegraphics[angle=0,scale=0.47]{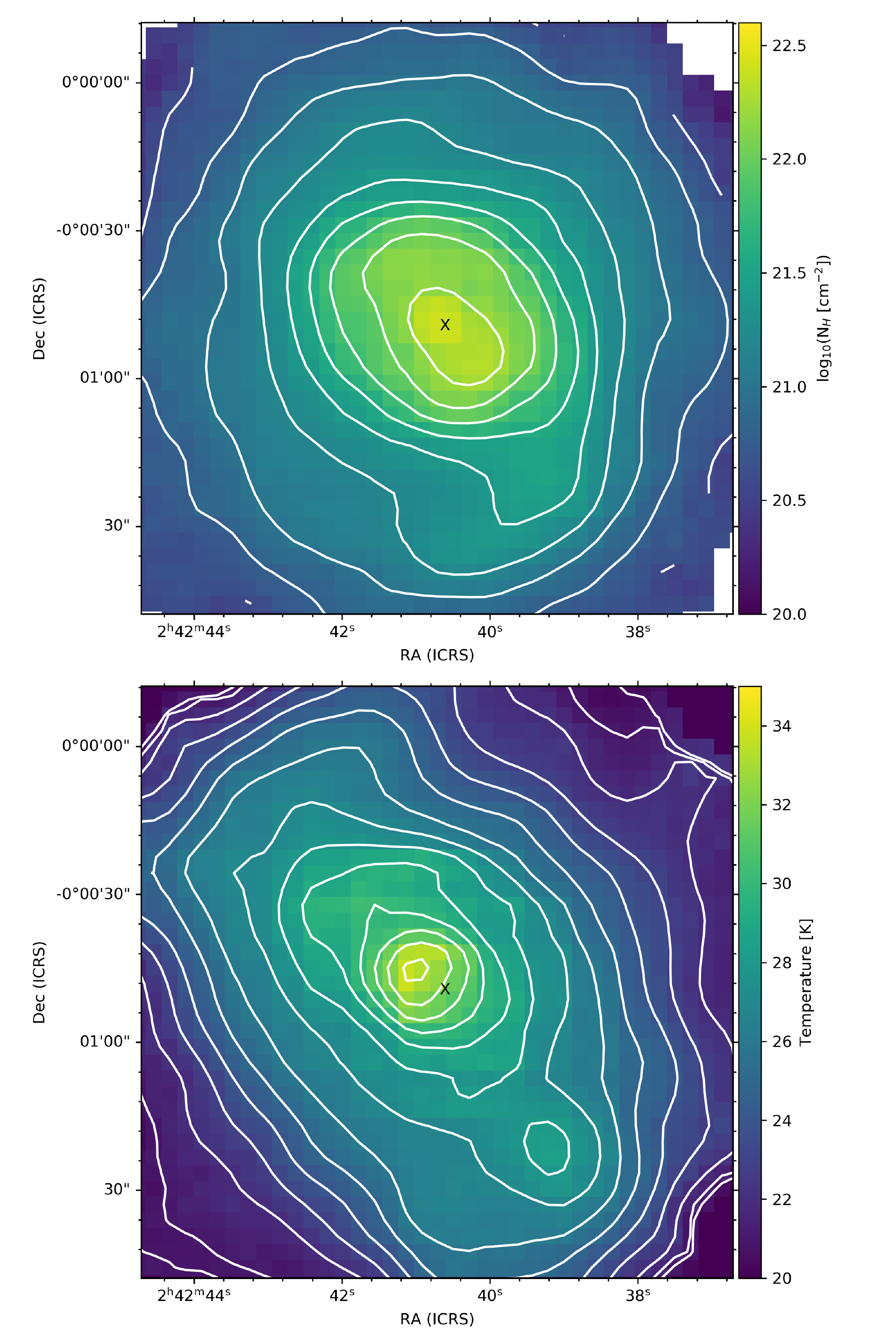}
\caption{ \textit{Top:} Hydrogen column density map, $N_{HI+H_{2}}$. Contours start at $\log{(N_{HI+H_{2}} (cm^{-2}))} = 20$ and increase in steps of $0.2$dex. \textit{Bottom:} Temperature map. Contours start at $20$K and increase in steps of $1$K. Black cross shows the location of the AGN. 
\label{fig:fig5}}
\epsscale{2.}
\end{figure}

To support the analysis of the following sections, we have constructed temperature and column density maps of NGC~1068. We have combined our $89$ \um\ HAWC+ observations with archival \textit{Herschel Space Observatory} \citep{Pilbratt2010} observations at $70$, $160$, and $250$ \um\ taken with the PACS \citep{Poglitsch2010} and SPIRE \citep{Griffin2010} instruments. We binned each observation to the pixel scale, $6$\arcsec, of the $160$ \um\ \textit{Herschel} data. Then we extracted the intensity values of each pixel associated with the same part of the sky at each wavelength. Finally, we fit a modified blackbody function assuming a dust emissivity of $\epsilon_{\lambda} \propto \lambda^{1.6}$ \citep[e.g.][]{Boselli2012} and with the temperature and optical depth, $\tau$, as free parameters.  We estimate the column density using the relation $N_{HI+H_{2}} = \tau / (k \mu m_{H})$, where $\tau$ is the optical depth at 250 $\mu$m from the best fit modified blackbody function at a given pixel, $k = 0.1$ cm$^{2}$ g$^{-1}$ \citep{H1983},  $\mu = 2.8$ is the mean molecular mass per H molecule \citep{K2008}, and $m_{H}$ is the hydrogen mass. Figure \ref{fig:fig5} shows the estimated temperature and column density distributions within the same FOV ($2\arcmin~\times~2\arcmin$, $7.8~\times~7.8$~kpc) as Fig. \ref{fig:fig1}. 

\begin{figure*}[ht!]
\includegraphics[angle=0,scale=0.9]{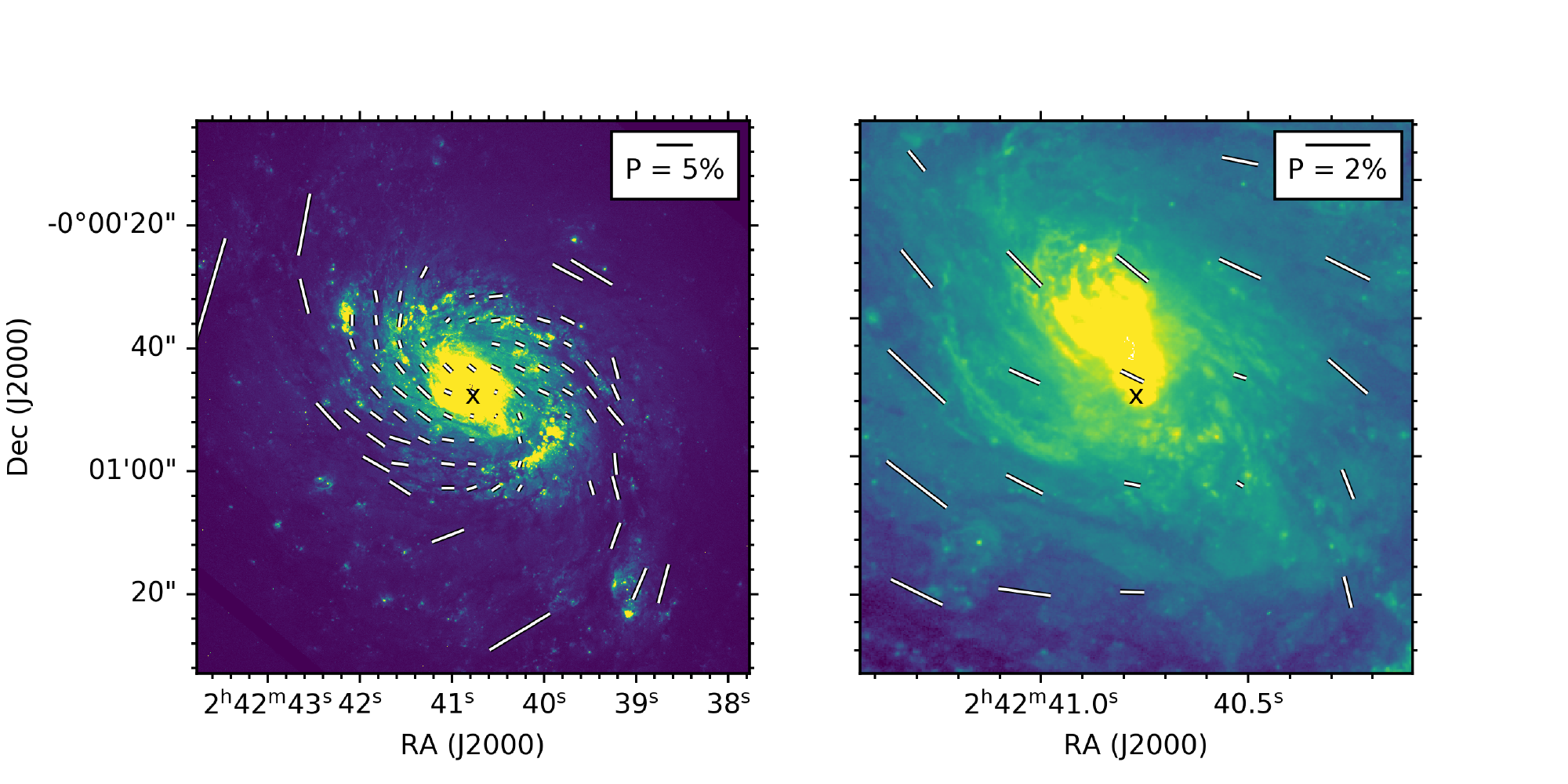}
\caption{\textit{Left:}  \textit{HST}/F555W image with the polarization vectors observed with HAWC+ within a $1.5\arcmin\ \times 1.5\arcmin$ ($5.9 \times 5.9$ kpc$^{2}$) FOV. \textit{Right:} Zoom-in of left image to show the central  $15\arcsec\ \times 15\arcsec$ ($0.98 \times 0.98$ kpc$^{2}$). The black cross shows the location of the active nucleus. Polarization vectors have been rotated by $90^{\circ}$ to show the inferred magnetic field.
\label{fig:fig6}}
\epsscale{2.}
\end{figure*}

The estimated dust temperatures lie in the range of $20-34$ K and peak in a ridge along the inner-bar. There is a strong peak temperature ($34\pm2$ K) displaced $\sim 7\arcsec$ ($\sim0.45$ kpc) NE from the position of the AGN (black cross). The dust temperature has another peak $\sim 45\arcsec$ ($\sim2.93$ kpc) SW spiral arm at the location of a large complex of HII regions and giant molecular clouds \citep[see Fig. \ref{fig:fig2}-right,][]{kaneko1989,tomoka2017}. Fig. \ref{fig:fig6} shows the location of the HII regions (bright compact sources) using the \textit{HST}/F555W ($\lambda_{c} = 0.54$ \um, $ \Delta\lambda = 0.11$ \um) filter to compare with our inferred magnetic field morphology. 

The column densities in the mapped area lie in the range $10^{20}-10^{22.4}$ cm$^{-2}$, corresponding to optical extinctions of A$_{V} \sim 0.05-14$ mag.  The distribution is extended along the inner-bar but is not as strongly peaked along the ridge as the temperatures. There is a weak maximum SW of the nucleus, and at the lowest contour levels there is an extension along the SW spiral arm. The morphology is consistent with HI absorption maps observed with the VLA with a beam size of $2\farcs1~\times~1\farcs0$ \citep{gallimore1994}. They found column densities in the range $[1-4]~\times~ 10^{21}$ cm$^{-2}$ in the region SW of the nucleus. Integrating over the column density map, we derive a mass of $[6.1~\pm~1.4]~\times~10^{8}$ M$_{\odot}$ in the central $2\arcmin~\times~2\arcmin$ ($7.8~\times~7.8$~kpc).

\subsection{On the role of magnetic fields in the spiral arms}\label{sec:BArms}

The main result of these observations is the detection of coherent magnetic fields over a $3$ kpc diameter region. Our column density map shows that all points have optical depths $< 0.2$ at $89$ \um. We converted the estimated hydrogen density, $N_{H}$, from Figure \ref{fig:fig5} to the visual extinction, $A_V$, using the standard ratio, $A_{V}/N_{H} = 5.35 \times 10^{-22}$ mag cm$^{-2}$ \citep{BSD1978}. Then, we used the typical extinction curve\footnote{The extinction curve used for the optical depth conversion can be found at \url{https://www.astro.princeton.edu/~draine/dust/dustmix.html}} of the Milky Way for R$_{V} = 3.1$ \citep{WD2001} to estimate the conversion $\tau_{0.55}/\tau_{89} = 4 \times 10^{-3}$ between optical depths at 0.55 \um~and 89 \um. Hence, at each line of sight  it is likely that we are sampling the integrated flux from aligned dust grains all the way through the galactic disk. In the outer regions of the map, the optical extinctions are also low to moderate, and there is a close alignment between visual tracers of spiral arms and our magnetic field vectors (see Fig. \ref{fig:fig1}, right and Fig. \ref{fig:fig6}, left). Within a diameter of $10$ kpc, CO is also aligned with the optical spiral structure, and H$_{\alpha}$ residual velocities reveal streaming motions along the arms \citep{D97}. 

While we discussed above some of the shortcomings of mapping magnetic fields with optical polarimetry, the V-band vectors of figure 2 in \citet{scar91} or the data shown by \citet{WT1987} line up well with our mathematical model. In the outer regions ($r >  15\arcsec$, outside the starburst ring) that is also true for our FIR polarimetry. Inside the starburst ring, where the optical depths are high (Fig. \ref{fig:fig5}-top), the V-band degree of polarization is very small, but in general, the directions follow our model. The FIR emission, on the other hand, is more highly polarized and there are significant deviations from the model (see Section \ref{sec:AGN}). From the above, we infer 1) that our FIR polarimetry results are, indeed, tracing spiral structure, 2) that, as expected, they seem to offer significant advantages over optical polarimetry for measuring the bulk properties of the ISM, particularly along lines of sight with high optical depth, and 3) that the field lines appear to follow the directions of gas flows along spiral arms, which may indicate that the arms have strong shocks otherwise the field would cross the arms at a given angle.

\begin{figure*}[ht!]
\includegraphics[angle=0,scale=0.6]{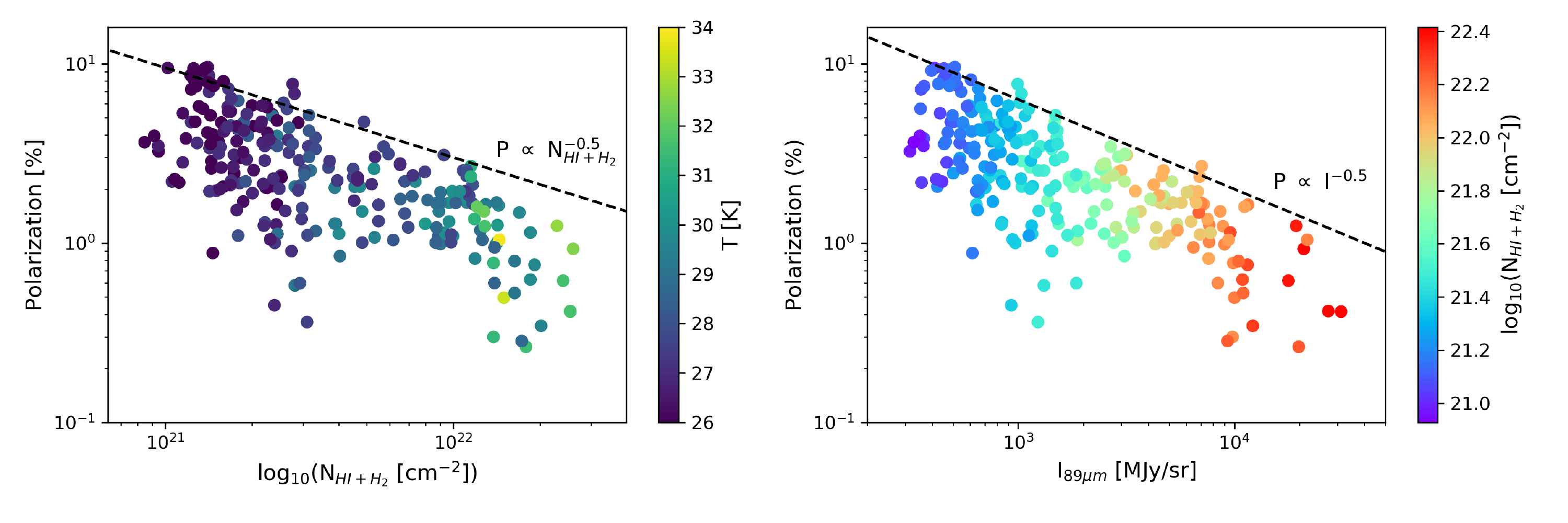}
\caption{Degree of polarization as a function of the column density (left) and surface brightness (right). Colorscale corresponds to the temperature (left) and column density (right). In all cases, polarization vectors with $\sigma_{P} < 5$\% are shown. A power-law (black dashed line) fit to the polarization vectors along the spiral arms are shown.
\label{fig:fig7}}
\epsscale{2.}
\end{figure*}

The magnetic field in the ISM is often modeled using a combination of constant and turbulent components. The trend of decreasing fractional polarization with increasing column density \citep{H99,jowh15, Jon15} provides an indirect measurement of the effect of the turbulent component. For maximally aligned dust grains along a line of sight with a constant magnetic field direction, the degree of polarization, $P$, in emission will be constant with optical depth $\tau$. If there is a region along the line of sight with some level of dust grain misalignment, it will result in a reduced degree of polarization. If the magnetic field direction in a region varies completely randomly along the line of sight with a well-defined scale length in $\tau$, we have $P \propto \tau^{-0.5}$ \citep{Jon15}. For any combination of constant and random components, the slope will be between these two limits \citep{JB15}. However, if 1) grain alignment decreases in denser regions, or 2) there is coherent cancellation of polarization (e.g. crossed field directions), or 3) there are regions of very high turbulence on very small scale lengths, the polarization can decline faster with optical depth than a slope of $-1/2$. 

Figure \ref{fig:fig7} shows log/log plots of polarization versus column density and surface brightness for all polarization vectors with $\sigma_{P} < 5$\%. The column density and dust temperature for each polarization vector is shown in color scale. It is clear that the polarization is higher when the temperature, column density, and fluxes are low, although the spatial anti-correlation of temperature and column density (Figure \ref{fig:fig5}) are the more fundamental relationships. This behavior has been previously observed in molecular clouds complexes in our Galaxy and Galactic center \citep[e.g.,][]{H99,C03}, which shows a depolarization effect as flux increases. The upper envelope of polarization measurements provides the least depolarized line-of-sights on the source, which allows to find the overall trend of the optical depth index. We use a linear fit to the log-log plot along the upper envelope of the measurements, which gives an optical depth index of $-0.5$ (black dashed line). This result indicates that most of the polarization measurements suffers of depolarization with an optical depth index steeper than $-0.5$, which may be due to high turbulence at small scales, effects of dust grain alignment towards dense regions and/or cross fields on the LOS.

The degree of polarization in our map is typically $< 3$\%, with a peak of $\sim 7$\% (with $P/\sigma_{P} \sim 3$). The polarization is systematically higher in the outer regions than in the starburst disk at radius $\le 25\arcsec$ ($1.6$ kpc). FIR polarimetric observations of Galactic clouds typically measure degree of polarization of $3-6$\% in the $60-100$ \um\ wavelength range \citep{dotson2000}. For our observations of NGC~1068, although we have an ordered polarization spiral pattern, we identify a potential disorder in the magnetic field on scales smaller than the resolution element, $8\arcsec$ ($520$ pc), of our observations. In a sense, this should not be surprising. The forces creating large-scale structures like spiral arms result from gravity and rotation. Where there are additional local sources of energy such at those arising from the assembly and disruption of star-forming complexes, there will be multiple additional mechanisms for distorting pre-existing magnetic fields or generating new ones with orientations that differ from the large-scale field. The larger the fraction of the ISM involved in these processes, the greater the average misalignment with the general field and the greater the depolarization of the emergent FIR radiation. This is consistent with studies that show that Galactic star-forming complexes with scale-lengths of tens of parsecs are misaligned with the Galaxy’s field on kiloparsec scales \citep{S2011}. One of the most significant aspects of our observations may be that in spite of the extremely high star formation rate in the starburst disk of NGC 1068, there is still a clearly discernible spiral field over much of its extent.

The alignment efficiency of dust grains may be an issue in very dense molecular cloud cores where the extinction exceeds A$_{V} \sim 20$ and no embedded Young Stellar Objects (YSOs) illuminate the dust  \citep{JB15}. To determine the effect of loss of grain alignment on our observations, we need to calculate what fraction of the flux in our 53 and 89 \um\ beams could arise from starless or pre-stellar molecular cores \citep{caselli2013}. Using $^{13}$CO and 1.1 mm dust continuum observations of the Milky Way, \citet{BH2014} find that only about 10\% or less of the mass in Giant Molecular Clouds (GMC) is contained in compact regions with A$_{v} > 10$ and temperatures of T$\sim10$ K. Given that some of these regions may contain a YSO that could produce enough radiation to align grains \citep[e.g.][]{jones2016}, this must be considered an upper limit to the mass fraction in starless cores. Given this upper limit on the mass fraction of GMCs in the form of starless cores and the $20-34$ K range of temperatures we derive for the dust emission in our beam (Figure \ref{fig:fig5}) for NGC~1068, the fraction of the flux in our beam from cloud cores with unaligned grains must be less than 1\%. Thus, in our 53 and 89~\um\ beams ($5-8$\arcsec), loss of grain alignment cannot be a significant contributing factor to the trend seen in Figure \ref{fig:fig7}. This leaves the coherent cancelation of polarization and/or regions of very high turbulence on small scale lengths as potential explanations for the slope in Figure \ref{fig:fig7}.
 
\subsection{The pitch angle and masses}\label{sec:PA}

As the density wave passes through the host galaxy, it enhances the star formation. The spiral density wave theory predicts that the pitch angle may vary with wavelength, as described in Section \ref{sec:int}. NGC~1068 has been studied across the electromagnetic spectrum where the pitch angle of the arms has been estimated to be $20.6\pm4.5^{\circ}$ at 0.46 \um\ \citep{Berrier2013}, $17.3 \pm 2.2^{\circ}$ using R band images with the 2.5-m Las Cumbres Observatory (LCO) \citep{Seigar2008}, $15^{\circ}$ using the $H_{\alpha}$ velocity field \citep{emsellen2006}, and $7-10^{\circ}$ using the CO (J=1-0) emission \citep{PSM1991}. We estimated a pitch angle of $16.9^{+2.7}_{-2.8}$$^{\circ}$ in the magnetic field inferred from 89 \um\ observations, which is between the pitch angles estimated in the spiral arms at 0.46 \um\ and the CO (J=1-0) emission. The material in the galaxy which is sampled by our FIR measurements is spatially coincident with the $H_{\alpha}$ velocity field \citep[see fig. 7 by][]{emsellen2006}. The analysis by \citet{DSW1998} computed that the $H_{\alpha}$ velocity field pattern is spatially correlated with the starbursting HII regions which was enhanced by {a burst in the past 30 Myr. Later, \citet{emsellen2006} found that the $H_{\alpha}$ and CO intensity distributions are not spatially coincident due to star formation and extinction. We conclude that the matter sampled by our estimated pitch angle follows the starbursting regions along the spiral arms.

The spiral density wave theory has also shown that the concentration of host galaxy mass and nuclear mass determines the pitch angle \citep{RRS1975}. As a consequence, recent studies have gone further and empirically shown that the spiral arm pitch angle, $\Psi$, and the central black hole mass, $M_{BH}$, may be correlated in the modal density wave  \citep[e.g.][]{Seigar2008, Berrier2013} by the form

\begin{equation}
\log{(\frac{M_{BH}}{M_{\odot}})} = (8.21 \pm 0.16) - (0.062 \pm 0.009)\Psi
\end{equation}
\noindent
\citep{Berrier2013}. Using our estimated pitch angle of $16.9^{+2.7}_{-2.8}$$^{\circ}$, we estimate a nuclear mass of $\log(M/M_{\odot})= 7.16^{+0.46}_{-0.51}$. Our result is in agreement, within the uncertainties, with the black hole mass of $\log(M_{BH}/M_{\odot})=6.95 \pm 0.02$ estimated using maser modeling on the central pc of NGC~1068 \citep{LB2003}.

\subsection{The central region of NGC~1068}\label{sec:AGN}

We found several features between the measured PA of polarization and the magnetic field model within $\sim1$ kpc of the nucleus. In the map of polarized flux (right-hand panel of Figure \ref{fig:fig1}), we identify: 

\begin{itemize}
\item a low polarization, $0.6\pm0.3$\%, within an $8\arcsec$ ($0.52$ kpc) diameter beam centered on the AGN, 
\item a strong peak in the polarized flux at PA~$\sim22^{\circ}$ at a radius of $\sim4\arcsec$ ($\sim0.26$ kpc) from the position of the AGN, 
\item three distinct minima at (a) at PA~$\sim22^{\circ}$, $8\arcsec~\le~r~\le~16\arcsec$ ($0.52-1$~kpc); (b) at PA~$\sim212^{\circ}$, $3\arcsec \le r \le 12\arcsec$ ($0.20-0.78$ kpc); and (c) at PA~$\sim240^{\circ}$, $14\arcsec \le r \le 18\arcsec$ ($0.91-1.17$ kpc), and 
\item significant discrepancies between our observed polarization directions and our spiral model.
\end{itemize}
  
The polarization of the nuclear region has been extensively studied from ultraviolet (UV) to MIR wavelengths. The UV-Optical polarization is very high, $~15$\% (intrinsic polarization), and it is attributed to dust scattering from the NE regions of the NLR within the central kpc \citep[e.g.][]{AM85,AHM94,K99}. The NIR polarization, $\sim7$\% (intrinsic polarization), is attributed to dichroic absorption of aligned dust grains in the pc-scale dusty obscuring torus around the active nucleus \citep[e.g.][]{B1988,Y1995,L1999,P1997,LR2015,G2015}. The MIR polarization, $<1$\%, is attributed to self-absorbed dust emission from aligned dust grains in the pc-scale optically thick dusty torus \citep{P07,LR2016}. We attribute our measured FIR polarization of $0.6\pm0.3$\% at the core due to depolarization from 1) beam, $\sim8$\arcsec, averaging of a more complex underlying field, 2) the high column density toward the core (Section \ref{sec:TNh}), and/or 3) turbulent environment caused by the jet.

It is interesting to note the difference between the nuclear polarization of the radio-quiet AGN NGC~1068 from the recently reported highly polarized radio-loud AGN, Cygnus A \citep{LR18}. At 89 \um, Cygnus A has a nuclear polarization of $\sim10$\% with a PA of polarization (B-vector) along the equatorial axis of the torus.  Although both galaxies are at different distances and their angular diameters in the plane of the sky are different, we can estimate the polarization of NGC 1068 at the same spatial scale of Cygnus A. We measure the polarization of NGC~1068 at the same physical scale ($8$ kpc diameter) of Cygnus A, and we obtain a polarization level of $\sim0.7$\% with a PA of polarization (E-vector) of $\sim144^{\circ}$, i.e. perpendicular to the overall extension of the polarized inner-bar (Fig. \ref{fig:fig1}). The core of NGC~1068 at 89 \um\ seems to be affected by depolarization arising from the inner-bar and/or an intrinsically weak magnetic field, while Cygnus A may have an intrinsically polarized core. Further magnetohydrodynamical models are required to test the influence of magnetic fields in AGN tori to compare between radio-loud and radio-quiet objects.

The peak of the polarized flux is located $\sim4$\arcsec\ ($\sim0.26$ kpc) NE from the AGN near the position of maximum dust temperature. This feature coincides with the northeastern radio lobe at 4.9 GHz \citep{WU1982}, which is associated with high-velocity OIII-emitting clouds within a bow-shock that separates the AGN outflow from the local ISM. \citet{YWS2001} observed a bright X-ray emission at the same position. The X-ray spectrum of this feature is a smooth continuum bremsstrahlung spectrum plus emission lines. Observations at $19.7-53$ \um\ show excess emission at $3\arcsec-6\arcsec$ ($0.20-0.39$ kpc) NE of the AGN after removal of an unresolved source at the position of the AGN \citep{LR2018a}. \citet{scar91} also found the region to be polarized at a level of $3$\%, which they attributed to the interaction of the jet with the ionization cone. At the angular resolution of the current FIR data, it is not possible to differentiate between the case in which the observed magnetic field at the position of peak polarized flux is generated by the bow shock or is intrinsic to the interstellar material with which the outflow is interacting.

The minima in polarized flux at $22^{\circ}$ (a) and $240^{\circ}$ (c) coincide with peaks in the $53$ \um\ map of \citet{LR2018a}, the locations of the most intense star formation in the starburst ring. The most likely explanation for the low polarization is that at these positions, the coherent large-scale field is completely dominated by randomly oriented fields generated during the collapse and destruction of a large number of star-forming cores.

The minimum in polarized flux at $212^{\circ}$ (b), in the direction of the counter-jet, may stem from a deficit in ISM density and CO intensity in that direction. The CO cloud density along the NE branch of the NIR bar and near the NE ionization cone appears to be greater than along the SW branch \citep[see, e.g.][]{Schinnerer2000,tosaki2017}.

The largest discrepancies between the measured polarization and the model of Section \ref{sec:Bfield} (see Figure \ref{fig:fig4}) lie at $5\arcsec \le r\le 15\arcsec$ at PA $\sim70^{\circ}$  and PA $\sim270^{\circ}$, in the directions of gaps in molecular cloud density along the starburst ring \citep{Schinnerer2000}. The gaps occur between the regions of most active star formation (see the $53$ \um\ map of \citet{LR2018a}) and the points at which the outer spiral arms diverge from the starburst ring (PA $\sim112^{\circ}$ and PA $\sim292^{\circ}$). The deviations of the observed field directions from the model are in the sense of a larger radial component in the observed vectors. In the sector toward PA $\sim 70^{\circ}$, the polarized flux is particularly strong (see Fig. \ref{fig:fig1}-right) and the vectors line up well with both the NE branch of the inner-bar (PA $\sim 45^{\circ}$) and visible dust lanes (see Fig. \ref{fig:fig2}-right). If the magnetic field traces the gas flow, this would be consistent with inward transport of gas at the leading edges of the inner-bar, with the highest present-day flow rates occurring along the NE branch of the bar. Magnetic fields tracing inward gas transport have been observed in the inner-bar of NGC 1097 in radio synchrotron polarization \citep{Beck2005}.


\section{Conclusions}\label{sec:CON}

We have presented the first FIR polarimetry of NGC 1068. We found a large-scale ($\sim3$ kpc) spiral pattern that we attribute to thermal emission from magnetically-aligned grains. There is a spatial and morphological correspondence between the 89 \um\ magnetic field vectors and other tracers, i.e. OIII, H$_{\alpha}$, optical, of spiral arms. We found a symmetric polarization pattern as a function of the azimuthal angle arising from the projection and inclination of the disk field component in the plane of the sky. We are able to explain this behavior with an axisymmetric spiral polarization pattern with pitch angle $16.9^{+2.7}_{-2.8}$$^{\circ}$, inclination $48.1^{+1.8}_{-1.9}$$^{\circ}$, and tilt angle $37.8 \pm 1.7^{\circ}$. The matter sampled by our estimated pitch angle follows the starbursting regions along the spiral arms, and, using the pitch angle-mass relationship, predicts a nuclear mass of $\log(M/M_{\odot}) = 7.16^{+0.46}_{-0.51}$.

Outside the starburst disk (radius $\le1.6$ kpc) in NGC 1068, the degree of polarization is similar to that seen in Milky Way sources. At smaller radii, it decreases with flux, dust temperature, and column density, and it has two minima near the locations of the ends of the NIR inner-bar in the starburst ring. This trend is consistent with dilution of the net spiral field by random fields created by injection of kinetic energy into the ISM by active star formation.

Inside the starburst ring ($<1.6$ kpc), we found evidence for large-scale coherent magnetic fields that align with visual tracers but not with our model field. The discrepancies are largest along the leading edges of the inner-bar. If the magnetic field traces gas flows, this is consistent with inward transport of gas induced by the bar. The intensity of polarized flux is stronger along position angles centered at $\sim 70^{\circ}$, suggesting that at the present time the strongest flows are along the NE branch of the inner-bar.

A peak in polarized flux intensity and dust temperature occurs at $\sim 4\arcsec$ ($\sim0.26$ kpc) NNE of the AGN near the location of the bow shock separating the AGN outflow from the surrounding ISM. This is consistent with the hypothesis that the magnetic field has been amplified at the shock interfaces along the edges of the outflow cavity, but our current angular resolution is insufficient to rule out the possibility that the field is intrinsic to the interstellar clouds with which the outflow is interacting. There is a minimum in polarized flux to the SSW of the AGN at the location of the counter-jet, possibly because of a current deficit of ISM in that direction.

The degree of FIR polarization at the position of the AGN is low, $1.3 \pm 0.3$\% within $5\arcsec$ ($0.33$ kpc) diameter at 53 \um\ and $0.6 \pm 0.3$\% within $8\arcsec$ ($0.52$ kpc) diameter at 89 \um. This is much lower than the $\sim10$\% FIR polarization of the radio-loud AGN in Cygnus A \citep{LR18}. The degree of polarization of NGC 1068 integrated over an $8$ kpc diameter region comparable to the physical scale of the region encompassed by the Cygnus A measurement is only $0.7$\%. 

The results presented here, along with our prior studies of M 82 and NGC 253 \citep{jones2019}, provide evidence that FIR polarimetry can be a valuable tool for studying magnetic field structure in external galaxies, particularly in regions of high optical depth.


\acknowledgments

We are grateful to the anonymous referee that helped to clarify and improve the text. Based on observations made with the NASA/DLR Stratospheric Observatory for Infrared Astronomy (SOFIA) under the GTO Program. SOFIA is jointly operated by the Universities Space Research Association, Inc. (USRA), under NASA contract NAS2-97001, and the Deutsches SOFIA Institut (DSI) under DLR contract 50 OK 0901 to the University of Stuttgart. Portions of this work were carried out at the Jet Propulsion Laboratory, operated by the California Institute of Technology under a contract with NASA. PACS has been developed by a consortium of institutes led by MPE (Germany) and including UVIE (Austria); KU Leuven, CSL, IMEC (Belgium); CEA, LAM (France); MPIA (Germany); INAF-IFSI/OAA/OAP/OAT, LENS, SISSA (Italy); IAC (Spain). This development has been supported by the funding agencies BMVIT (Austria), ESA-PRODEX (Belgium), CEA/CNES (France), DLR (Germany), ASI/INAF (Italy), and CICYT/MCYT (Spain). SPIRE has been developed by a consortium of institutes led by Cardiff University (UK) and including Univ. Lethbridge (Canada); NAOC (China); CEA, LAM (France); IFSI, Univ. Padua (Italy); IAC (Spain); Stockholm Observatory (Sweden); Imperial College London, RAL, UCL-MSSL, UKATC, Univ. Sussex (UK); and Caltech, JPL, NHSC, Univ. Colorado (USA). This development has been supported by national funding agencies: CSA (Canada); NAOC (China); CEA, CNES, CNRS (France); ASI (Italy); MCINN (Spain); SNSB (Sweden); STFC, UKSA (UK); and NASA (USA).

%

\vspace{5mm}
\facilities{SOFIA (HAWC+), \textit{Herschel} (PACS, SPIRE), \textit{HST} (WFPC2, ACS)}


\software{\textsc{astropy} \citep{2013A&A...558A..33A}, 
\textsc{PyMC3} \citep{pymc},
\textsc{aplpy} \citep{RB2012},
\textsc{matplotlib} \citep{hunter2007}. 
          }


\begin{thebibliography}{}
\bibitem[Andersson, Lazarian \& Vaillancourt(2015)]{ALV2015} Andersson, B.-G., Lazarian, A., Vaillancourt, J. E. 2015, ARA\&A, 53, 501.
\bibitem[Antonucci \& Miller(1985)]{AM85} Antonucci, R. R. J., Miller, J. S. 1985, ApJ, 297, 621
\bibitem[Antonucci, Hurt, \& Miller(1994)]{AHM94} Antonucci, R. R. J., Hurt, T., Miller, J. S. 1994, ApJ, 430, 210
\bibitem[Astropy Collaboration et al.(2013)]{2013A&A...558A..33A} Astropy Collaboration, Robitaille, T.~P., Tollerud, E.~J., et al.\ 2013, \aap, 558, A33 
\bibitem[Athanassoula(1984)]{Athanassoula1984}Athanassoula E., 1984, Phys. Rep., 114, 319


\bibitem[Bailey et al.(1988)]{B1988} Bailey, J., Axon, D. J., Hough, J. H., Ward, M. J., McLean, I., Heathcote, S. R. 1988, MNRAS, 234, 899
\bibitem[Balick \& Heckman(1985)]{BH1985} Balick, B., Heckman, T. 1985, ApJ, 90, 197
\bibitem[Battisti \& Heyer(2014)]{BH2014} Battisti, A. J., \& Heyer, M. H. 2014, ApJ, 780, 173
\bibitem[Beck, Klein \& Wielebinski(1987)]{BKW1987} Beck, R., Klein, U., Wielebinski, R. 1987, A\&A, 186, 95
\bibitem[Beck \& Gaensler(2004)]{BG2004} Beck, R., Gaensler, B. M. 2004, NewAR, 48, 1289
\bibitem[Beck(2015)]{beck2015} Beck, R. 2015, A\&ARv, 24, 4
\bibitem[Beck et al.(2005)]{Beck2005} Beck, R., Fletcher, A., Shukurov, A., Snodin, A., Sokoloff, D. D., Ehle, M., Moss, D., Shoutenkov, V. 2005, A\&A, 444, 739
\bibitem[Berrier et al.(2013)]{Berrier2013}  Berrier, J. C., Davis, B. L., Kennefick, D., Kennefick, J. D., Seigar, M. S., Barrows, R. S., Hartley, M., Shields, D., Bentz, M. C., Lacy, C. H. S. 2013, ApJ, 769, 132
\bibitem[Bohlin, Savage \& Drake(1978)]{BSD1978} Bohlin, R. S., Savage, B. D., Drake, J. F. 1978, ApJ, 224, 132
\bibitem[Boselli et al.(2012)]{Boselli2012} Boselli, A., Ciesla, L., Cortese, L., et al. 2012, A\&A, 540, A54
\bibitem[Braun, Heald \& Beck(2010)]{BHB2010} Braun, R., Heald, G., Beck, R. 2010, A\&A, 514, 42
\bibitem[Brinks et al.(1997)]{brinks1997} Brinks, E., Skillman, E. D., Terlevich, R. J., Terlevich, E. Ap\&SS, 248, 23

\bibitem[Cabral \& Leedom(1993)]{CL1993} Cabral, B., \& Leedom, L. C. 1993, in Proceedings of the 20th annual conference on Computer graphics and interactive techniques, ACM, 263–270
\bibitem[Caselli(2013)]{caselli2013} Caselli, P. 2013, New Trends in Radio Astronomy in the ALMA Era: The 30th Anniversary of Nobeyama Radio Observatory, 476, 169
\bibitem[Chuss et al.(2003)]{C03} Chuss, D. T., Davidson, J. A., Dotson, J, L., Dowell, C. D., Hildebrand, R. H., Novak, G., Vaillancourt, J. E. 2003, ApJ, 599, 1116

\bibitem[Das et al.(2006)]{D2006} Das, V., Creenshaw, D. M., Kraemer, S. B., Deo. T. P. 2006, AJ, 132, 620
\bibitem[Davies, Sugai \& Ward(1998)]{DSW1998} Davies, R. I., Sugai, H., Ward, M. J. 1998, MNRAS, 300, 388
\bibitem[de Vaucouleurs et al.(1991)]{deVaucouleurs1991} de Vaucouleurs, G., de Vaucouleurs, A., Corwin, H. G., Jr., Buta, R. J., Paturel, G., Fouque, P. 1991, Third Reference Catalogue of Bright Galaxies, Springer, New York, NY (USA).
\bibitem[Dehnen et al.(1997)]{D97} Dehnen, W., Bland-Hawthorn, J., Quirrenbach, A., Cecil, G, N. 1997, Ap\&SS, 248, 33
\bibitem[Dobbs \& Baba(2014)]{DB2014} Dobbs, C., Baba, J. 2014, PASP, 31, 35
\bibitem[Dotson et al.(2000)]{dotson2000} Dotson, J. L., Davidson, J., Dowell, C. D., Schleuning, D. A., Hildebrand, R. H. 2000, ApJS, 128, 335
\bibitem[Dowell(1997)]{dowell1997} Dowell, C. D. 1997, ApJ,  487, 237
\bibitem[Dowell et al.(2010)]{Dowell2010} Dowell, C. D., Cook, B. T., Harper, D. A., et al. 2010, Proc. SPIE, 7735, 77356H

\bibitem[Emsellem et al.(2006)]{emsellen2006} Emsellem, E.  Fathi, K., Wozniak, H., Ferruit, P., Mundell, C. G., Schinnerer, E. 2006, 365, 367MNRAS, 

\bibitem[Fendt, Beck \& Neininger(1998)]{FBN1998} Fendt, Ch., Beck, R., Neininger, N. 1998, A\&A, 335, 123
\bibitem[Fletcher et al.(2011)]{flecher2011} Fletcher, A., Beck, R., Shukurov, A., Berkhuijsen, E. M., Horellou, C. 2011, MNRAS, 412, 2396

\bibitem[Gallimore et al.(1994)]{gallimore1994} Gallimore, J. F., Baum, S. A., O'Dea, C. P., Brinks, E., Pedlar, A. 1994, ApJ, 422, 13
\bibitem[Gratadour et al.(2015)]{G2015} Gratadour, D., Rouan, D., Grosset, L., Boccaletti, A., Cl{\'e}net, Y.\ 2015, \aap, 581, L8.
\bibitem[Griffin et al.(2010)]{Griffin2010} Griffin, M. J., Abergel, A., Abreu, A., et al. 2010, A\&A, 518, L3

\bibitem[Harper et al.(2018)]{Harper2018} Harper, D. A., Runyan, M. C., Dowell, D. A., et al. 2018, JAI, 740008H
\bibitem[Hildebrand(1983)]{H1983} Hildebrand, R. H., 1983, QJRAS, 24, 267 
\bibitem[Hildebrand et al.(1999)]{H99} Hildebrand, R. H., Dotson, J. L., Dowell, C. D., Schleuning, D. A., Vaillancourt, J. E. 1999, ApJ, 516, 834
\bibitem[Hunter(2007)]{hunter2007} Hunter, J. D. 2007, Computing in Science \& Engineering, 9, 3.

\bibitem[Ichikawa, Okamura \& Kaneko(1987)]{IOK1987} Ichikawa, S.-I., Okamura, S., Kaneko, N. 1987, PASJ, 39, 411

\bibitem[Jones \& Whittet(2015)]{jowh15} Jones, T.~J., Whittet, D., C.~B.\ 2015, Polarimetry of Stars and Planetary Systems, 147 
\bibitem[Jones(2015)]{Jon15} Jones, T.~J.\ 2015, Magnetic Fields in Diffuse Media, 407, 153 
\bibitem[Jones et al.(2015)]{JB15} Jones, T.~J., Bagley, M., Krejny, M., Andersson, B.-G.,  Bastien, P.\ 2015, \aj, 149, 31 
\bibitem[Jones et al.(2016)]{jones2016} Jones, T. J., Gordon, M., Shenoy, D., Gehrz, R. D., Vaillancourt, J. E., Krejny, M. 2016, AJ, 151, 156
\bibitem[Jones et al.(2019)]{jones2019} Jones, T. J., Dowell, C. D., Lopez-Rodriguez, E., Zweibel, E. G., Berthoud, M., Chuss, D. T., Goldsmith, P. F., Hamilton, R. T., Hanany, S., Harper, D. A., Lazarian, A., Looney, L. W., Michail, J. M., Morris, M. R., Novak, G., Santos, F. P., Sheth, K., Stacey, G. J., Staguhn, J., Stephens, I. W., Tassis, K., Trinh, C. Q., Volpert, C. G., Werner, M., Wollack, E. J., HAWC+ Science Team. 2019, ApJ, 870, 


\bibitem[Kaneko et al.(1989)]{kaneko1989}  Kaneko, N., Morita, K., Fukui, Y., Sugitani, K., Iwata, T., Nakai, N., Kaifu, N., Liszt, H. S. 1989, ApJ, 337, 691.
\bibitem[Kauffmann et al.(2008)]{K2008} Kauffmann, J., Bertoldi, F., Bourke, T. L., Evans II, N. J., Lee, C. W., 2008, A\&A, 993, 1017
\bibitem[Keel \& Weedman(1978)]{KW1978} Keel, W. C., Weedman, D. W., 1978, AJ, 83, 1
\bibitem[Kishimoto(1999)]{K99} Kishimoto, M. 1999, ApJ, 518, 676
\bibitem[Kronberg(1994)]{kron1994} Kronberg, P. P. 1994, Reports on Progress in Physics, Volume 57, Issue 4.

\bibitem[Lin \& Shu(1964)]{LS1964} Lin, C. C., Shu, F. H. 1964, ApJ, 140, 646
\bibitem[Lindblad(1960)]{lindblad1960} Lindblad, P. O. 1960, StoAn, 21, 3.
\bibitem[Lodato \& Bertin(2003)]{LB2003} Lodato, G, Bertin, G. 2003, A\&A, 398, 517
\bibitem[Lopez-Rodriguez et al.(2015)]{LR2015} Lopez-Rodriguez, E., Packham, C., Jones, T. J., Nikutta, R., McMaster, L., Mason, R. E., Elvis, M., Shenoy, D., Alonso-Herrero, A., Ram\'irez, E., Gonz\'alez Mart\'in, O., H\"onig, S. F., Levenson, N. A., Ramos Almeida, C., Perlman, E. 2015, MNRAS, 452, 1902
\bibitem[Lopez-Rodriguez et al.(2016)]{LR2016} Lopez-Rodriguez, E., Packham, C., Roche, P. F., Alonso-Herrero, A., Diaz-Santos, T., Nikutta, R., Gonzalez-Martin, O., Alvarez, C. A., Esquej, P., Rodriguez Espinosa, J. M., Perlman, E., Ramos Almeida, C., Telesco, C. M. 2016, MNRAS, 458, 3851
\bibitem[Lopez-Rodriguez et al.(2018a)]{LR2018a} Lopez-Rodriguez, E., Fuller, L., Alonso-Herrero, A., Efstathiou, A., Ichikawa, K., Levenson, N. A., Packham, C., Radomski, J., Ramos Almeida, C., Benford, D. J., Berthoud, M., Hamilton, R., Harper, D., Kov\'avcs, A., Santos, F.~P., Staguhn, J., Herter, T. 2018a, ApJ, 859, 99
\bibitem[Lopez-Rodriguez et al.(2018b)]{LR18}  Lopez-Rodriguez, E., Antonucci, R. R. J., Chary, R.-R., Kishimoto, M. 2018b, ApJL, 861, 23
\bibitem[Lumsden et al.(1999)]{L1999} Lumsden, S. L., Moore, T. J. T., Smith, C., Fujiyoshi, T., Bland-Hawthorn, J., Ward, M. J. 1999, MNRAS, 303, 209


\bibitem[Packham et al.(1997)]{P1997} Packham, C., Young, S., Hough, J. H., Axon, D. J., Bailey, J. A. 1997, MNRAS, 288, 375
\bibitem[Packham et al.(2007)]{P07} Packham, C., Young, S., Fisher, S., Volk, K., Mason, R., Hough, J. H., Roche, P. F., Elitzur, M., Radomski, J., Perlman, E. 2007, ApJ, 661, 29
\bibitem[Pavel \& Clemens(2012)]{pave12} Pavel, M.~D., Clemens, D.~P.\ 2012, \apjl, 761, L28 
\bibitem[Pilbratt et al.(2010)]{Pilbratt2010} Pilbratt, G. L., Riedinger, J. R., Passvogel, T., et al. 2010, A\&A, 518, L1
\bibitem[Planck Collaboration VI(2018)]{Plank2018} Planck Collaboration, et al. 2018,  arXiv:1807.06209v1
\bibitem[Planesas, Scoville \& Myers(1991)]{PSM1991} Planesas, P., Scoville, N., Myers, S. T. 1991, ApJ, 369, 364
\bibitem[Poglitsch et al.(2010)]{Poglitsch2010} Poglitsch, A., Waelkens, C., Geis, N., et al. 2010, A\&A, 518, L2
\bibitem[Pour-Imani et al.(2016)]{PI2016} Pour-Imani, H., Kennefick, D., Kennefick, J., Davis, B. L., Shields, D. W., Shameer Abdeen, M. 2016, ApJ, 827, 2

\bibitem[Roberts, Roberts \& Shu(1975)]{RRS1975} Roberts, W. W., Jr., Roberts, M. S., Shu, F. H. 1975, ApJ, 196, 381
\bibitem[Robitaille \& Bressert(2012)]{RB2012} Robitaille, T., Bressert, E. 2012, Astrophysics Source Code Library (ASCL), 08017.
\bibitem[Ruiz-Granados et al.(2010)]{RG2010} Ruiz-Granados, B., Rubi{\~n}o-Mart{\'{\i}}n, J.~A. Battaner, E. 2010, A\&A, 522, 73

\bibitem[Salvatier, Wiecki \& Fonnesbeck(2016)]{pymc} Salvatier J., Wiecki T.V., Fonnesbeck C. (2016) Probabilistic programming in Python using PyMC3. PeerJ Computer Science 2:e55
\bibitem[Scarrott et al.(1987)]{scar87} Scarrott, S.~M., Ward-Thompson, D., Warren-Smith, R.~F.\ 1987, \mnras, 224, 299 
\bibitem[Scarrott et al.(1991)]{scar91} Scarrott, S.~M., Rolph, C.~D., Wolstencroft, R.~W., Tadhunter, C.~N.\ 1991, \mnras, 249, 16P 
\bibitem[Schinnerer et al.(2000)]{Schinnerer2000} Schinnerer, E., Eckart, A., Tacconi, L. J., Genzel, R., Downes, D.  2000, ApJ, 533, 850
\bibitem[Scoville et al.(1988)]{scoville1988}Scoville N. Z., Matthews K., Carico D. P., Sanders D. B., 1988, ApJ, 327, L61
\bibitem[Seigar(2008)]{Seigar2008} 	Seigar, M.~S., Kennefick, D., Kennefick, J., Lacy, C. H. S. 2008, ApJ, 678, 93
\bibitem[Shu(2016)]{shu2016} Shu, F. H. 2016, ARA\&A, 54, 667
\bibitem[Stephens et al.(2011)]{S2011} Stephens, I. W., Looney, L. W., Dowell, C. D., Vaillancourt, J. E., Tassis, K. 2011, ApJ, 728, 99

\bibitem[Tanaka, Yagi \& Taniguchi(2017)]{TYT2017} Tanaka, I., Yago, M., Taniguchi, Y. 2017, PASJ, 69, 90
\bibitem[Telesco \& Harper(1980)]{TH1980} Telesco, C. M., Harper, D. A. 1980. ApJ, 235, 392
\bibitem[Telesco et al.(1984)]{telesco1984} Telesco, C. M., Becklin, E. E., Wynn-Williams, C. G., Harper, D. A. 1984, ApJ, 282, 427
\bibitem[Telesco \& Decher(1988)]{TD1988} Telesco, C. M., Decher, R. 1988, ApJ, 334, 573
\bibitem[Tomoka et al. (2017)]{tomoka2017} Tomoka, T., Kotaro, K., Nanase, H., Kunihiko, T., Fumi, E., Takuma, I., Shuro, T., Taku, N., Akio, T., Yoichi, T. 2017, PASJ, 69, 18
\bibitem[Tosaki et al.(2017)]{tosaki2017} Tosaki, T., Kohno, K., Harada, N., Tanaka, K., Egusa, F., Izumi, T., Takano, S., Nakajima, T., Taniguchi, A., Tamura, Y. 2017, PASJ, 69, 18
\bibitem[Thronson et al.(1989)]{thronson1989} Thronson, H. A. Jr., Hereld, M., Majewski, S., Greenhouse, M., Johnson, P., Spillar, E., Woodard, C. E., Harper, D. A., Rauscher, B. J. 1989, ApJ, 343, 158 


\bibitem[Vaillancourt et al.(2007)]{Vaillancourt2007}Vaillancourt, J. E., Chuss, D. T., Crutcher, R. M., et al. 2007, Proc. SPIE, 6678, 66780D

\bibitem[Ward-Thompson(1987)]{WT1987} Ward-Thompson D., 1987, PhD Thesis, University of Durham, `A polarisation study of spiral galaxies'
\bibitem[Wardle \& Kronberg(1974)]{WK1974} Wardle, J.~F.~C., Kronberg, P.~P. 1974, ApJ, 194, 249 
\bibitem[Weedman(1985)]{weedman1985} Weedman, D. W. 1985, ApJS, 57, 523
\bibitem[Weingartner \& Draine(2001)]{WD2001} Weingartner, J. C., Draine, B. T. 2001, ApJ, 548, 296
\bibitem[Wilson \& Ulvestad(1982)]{WU1982} Wilson, A. S., Ulvestad, J. S. 1982, ApJ, 263, 576

\bibitem[Young et al.(1995)]{Y1995} Young, S., Hough, J. H., Axon, D. J., Bailey, J. A., Ward, M. J. 1995, MNRAS, 272, 513
\bibitem[Young, Wilson \& Shopbell(2001)]{YWS2001} Young, A. J., Wilson, A. S., Shopbell, P. L. 2001, ApJ, 556, 6

\bibitem[Zweibel \& Heiles(1997)]{ZH1997} Zweibel, E. G., Heiles, C. 1997, Nature, 385, 131

\end{thebibliography}
\end{document}